\documentclass[10pt,journal,compsoc]{IEEEtran}

\ifCLASSOPTIONcompsoc
  \usepackage[nocompress]{cite}
\else
  \usepackage{cite}
\fi

\ifCLASSINFOpdf
\else
\fi

\begin{document}
\title{How Relevant is the Turing Test\\ in the Age of Sophisbots?\\}

\author{Dan Boneh, Stanford University \hfill dabo@cs.stanford.edu\\
	Andrew J. Grotto, Stanford University \hfill grotto@stanford.edu\\
	Patrick McDaniel, Pennsylvania State University \hfill mcdaniel@cse.psu.edu \\
	Nicolas Papernot, University of Toronto \hfill nicolas.papernot@utoronto.ca}

\IEEEtitleabstractindextext{%
\begin{abstract}
Popular culture has contemplated societies of thinking machines for generations, envisioning futures from utopian to dystopian. These futures are, arguably, here now---we find ourselves at the doorstep of technology that can at least simulate the appearance of thinking, acting, and feeling.  The real question is: now what?
\end{abstract}

}

\maketitle

\IEEEdisplaynontitleabstractindextext

\IEEEpeerreviewmaketitle

\IEEEPARstart{P}{opular} culture has contemplated societies of thinking machines for generations, envisioning futures from utopian to dystopian. Visions of HAL and R2D2 have excited our imagination and at the same time prompted concerns about how these new entities will impact us and our society at large. These futures are, arguably, here now---we find ourselves at the doorstep of technology that can at least simulate the appearance of thinking, acting, and feeling. The real question is: now what?

Was this the future that Alan Turing famously envisioned in 1950 when he created the Turing Test to tell machines apart from human based on whether a machine could fool a human that is interacting with it into thinking it is human?  Maybe and maybe not. Advanced uses of ML and AI allow us to engineer programs that are capable of outputting hyper-realistic communication, images or even potentially full human personas. While there are numerous positive uses of such technology, does it have darker implications if used with the intent to cause harm? 

We take the example of sophisticated online bots able to emulate human behavior and interact with us seamlessly. As they make the distinction between authentic and fake communication harder than it is already today, these sophisbots could have a profound impact on society by gradually manipulating our trust in content like images or videos. We argue why technical solutions, while important, should be complemented with efforts involving informed policy and international norms to accompany these technological developments. Indeed, we believe it is essential to foster increased civic literacy of the nature of one’s interactions with systems deploying ML to ensure a positive impact of the advances they enable.

\section*{Do machines capable of passing the Turing test exist today?}

Machine learning can now emulate human behavior, thought processes, and strategies, to the point of human indistinguishability between humans and machines in certain contexts. Google's Duplex system makes reservations by conversing with humans over the phone. Here, learning algorithms captured subtle artifacts of spoken English to replicate human artifacts such as hesitations or pauses, thereby generating speech that is very conversational and lifelike. In another domain, Christie's announced this past summer that it was the first auction house to sell art generated by a neural network. This questions the necessary involvement of human forms of creativity as a prerequisite to producing art that is enjoyable to humans. 

It is not hyperbole to suggest that such demonstrations, that were until now exclusively associated with humans, are only the first of many more to come. Machine learning techniques are no longer limited to responding to inputs created and curated by humans---like a request to place a reservation. They are now capable of producing original synthetic content completely from scratch using a class of machine learning algorithms known as ``generative''. This area of research has in many ways been revolutionized by a particular technique called GANs, which stands for Generative Adversarial Networks~\cite{GPM14}. GANs enable many creative applications beneficial to society: they range from automatically designing dental crowns that fit individual patients correctly to augmenting the creativity of artists by synthesizing musical notes. 

The GAN framework involves two machine learning models trained in a competing fashion. The first, called the generator, learns how to produce synthetic inputs, e.g., images or videos. The second, referred to as the discriminator, learns how to classify inputs as being real or ``fake'' (i.e., data output by the generator). This may remind the reader of the Turing test. Indeed, the discriminator is orchestrating some form of weaker but automated imitation game~\cite{T50}. Assuming neither model wins this game early in training, the discriminator gets better at telling apart synthetic and real inputs while the generator gradually gets better at producing synthetic inputs that the discriminator finds harder to detect. We revisit this relationship between the GAN framework and the imitation game later, since it has implications for the (lack of) existence of automated techniques for detecting machine-generated content.

It is exciting that scientific progress enables us to get closer to building machines that may pass the Turing test. The ability to generate synthetic media like videos that appear natural to humans has many positive applications: for instance, the episode IX of Star Wars is set to include the late actor Carrie Fisher. However, like with any technology, scientific progress made by researchers interested in generative machine learning could enable potentially less honorable applications by individuals with malicious intentions to bully, blackmail, extort, defame, or mislead. 

Take the example of research conducted by Suwajanakorn et al., in which they applied another type of generative ML model called a recurrent neural network to synthesize the face and voice of Barack Obama~\cite{SSK17}. The researchers were able to produce realistic video footage of the former US president giving an address from a text transcript of their choice. Although a careful human observer will notice some minor inconsistencies in the videos produced by the machine, it is clear that such a technique could be leveraged by malicious individuals to increase the sophistication of fake news spread by social media bots. This was demonstrated, most likely without malicious intentions, when a Belgian party created in May 2018 a video of Donald Trump urging Belgium to follow the US and withdraw from the Paris climate agreement. By circulating this video on social networks, the party's intent was to generate attention and debate around the issue of climate change. While the intention is likely not to deceive here, releasing this video contributes to eroding confidence from the general public in media. In another example, Yao et al. showed that machine learning is capable of generating fake Yelp reviews that humans not only confounded with real reviews but also found just as useful~\cite{YVC17}. 

In some way, this technology is close from passing a Turing test, albeit a passive one where the human consumes the machine's outputs but is unable to query it with specific inputs. 

\section*{What are the implications of ML emulating human behavior?}

All of these discoveries and observations lead to a single conclusion---we are rapidly reaching a point where computational algorithms can create nearly any form of human communication that is for all intents and purposes indistinguishable from reality. Apart from the many obvious positive uses, such as telepresence or human-computer interaction, what does this mean to us as a society and what does it do to online (and offline) public discourse, politics, due process, commerce, and society at large?

In the presence of individuals with malicious intent but low technical sophistication, the issue is first one of forgery. There is a history of written words and reported discourse being forged. Forged letters claimed in 1777 that General Washington thought the revolutionary war against Great Britain was a mistake. After multiple decades of improvements to the software created by Adobe, the expression ``to photoshop'' is now a synonym for digitally altering images. In short, the barrier to entry for manipulating content has been decreasing for centuries. Progress in ML is simply accelerating this process. Things that were less forgeable in the past---such as voice or video when compared to written words or photos---now are or soon will be.  

We already have a substantial problem with forged and out-of-context content---the Internet is plagued by fake reviews, fake posts, and fake people fueling misinformation.  We now know that carefully crafted social media messages were created to disrupt public discourse and influence opinion prior to the 2016 US presidential election. Suppose adversaries could automate this process and create such content algorithmically, an hypothesis reinforced in the controversial way researchers presented their recent results~\cite{RWC19} on natural language processing to the public.  How does that capability change the Internet and affect our society?

As the sophistication of adversaries increases, consider the hypothetical but logical conclusion of content forgery---a fake human we will refer to as a sophisbot.  Here, the sophisbot is a program running in the ether of social media or other infrastructure.  It never sleeps nor ages, and is not bound by geography, culture or conscience. It can have opinions, agendas, biases and can consume enormous amounts of information and maintain nearly infinite number of simultaneous conversations. Sophisbots can have real visages and personalities that will draw people to them.

Now take that same bot and give them a task.  Such tasks can be political, such as promoting one candidate over others, sowing mistrust among citizens, spreading racist or misogynistic propaganda, etc. Other tasks can be more on the human-scale, such as harassing an at-risk teenager into suicide, ruining personal relationships, or inducing a victim into doing something financially or personally risky. Perhaps less damaging but still raising ethical questions, the bots could simply promote products or services. Sophisbots would tirelessly use all the tools of forgery and social engineering to achieve their goals. Older technologies previously transformed communication in similar ways: email decreased the resources needed to craft sophisticated junk mail that could reach millions of individuals. Now consider that you can use machine learning to create billions of sophisbots simulating fake humans in a matter of seconds.

Such bots are some distance from those we see today. Nevertheless, Twitter-bots already generate fake-but-widely-read forged content and retweet others. A 2017 study by the Pew Research Institute shows that over half of the links on Twitter were posted by automated accounts. This reality was also observed in earlier academic studies~\cite{CGW10} or during the DARPA challenge on Twitter bots.  

In short, while sophisbots would not introduce a new problem, they would reinforce the existing issue of content forgery by potentially providing scalability to existing techniques for manipulating content. Adversaries have already exploited existing online services driven by automated reasoning for such malicious ends.  In one well publicized example, Microsoft's Tay chatbot was an online chat bot that learned to speak (in part) from user queries.  Adversaries on the Internet quickly learned of the service and trained it to post racist and inflammatory tweets~\cite{NN16}.  The service was shutdown within 24 hours of its launch. 

The real question is how do we identify and possibly eliminate these malicious bots and content from public discourse?

\section*{Do technological defenses exist?}

The central question is how do we defend against this form of malicious activity. There are really two answers to this question: science and policy. We first discuss the former.

\textbf{We argue as a matter of science, machine learning technology must evolve to make the systems and models accountable and its inputs and outputs reliably identifiable}~\cite{AC18}.  Digital forensic techniques are actively being developed to detect manipulated content. For instance, specialists commonly use the lack of camera-induced imperfections in synthetic images and videos to identify them. Unfortunately, the approach does not scale given the shortage of human experts; unless their work can be in large parts automated by technological solutions that identify content produced or edited by machines. We explore three such potential approaches, but stress that despite the partial progress they afford, all three are quite limited. Our analysis thus suggests that it is likely that there is no robust technological defense against this problem.

\subsubsection*{Detecting artifacts of synthetic content} A natural first approach is to automate the process of digital forensics and attempt to identify machine-manipulated content by detecting its imperfections. For instance, techniques for manipulating videos often introduce specific video imperfections that can be detected. For instance, generators that rely on deep learning to produce fake videos known as deepfakes often operate by face-swapping.  Body movements and proportions are typically unchanged from the stand-in actor. Techniques such as Eulerian Video Magnification could help to identify human pulse in videos~\cite{EHD15}. In principle, a detector could identify a deepfake by detecting these and other imperfections in the video. In fact, the Defense Advanced Research Projects Agency (DARPA) is running a media forensics program called MediFor that is funding research to develop robust detectors that operate this way. Researchers are developing tools to detect deepfakes through physiological inconsistencies, such as detecting an irregular eye blinking pattern, or lack of blinking altogether~\cite{LCL18}. To facilitate learning detectors for synthetic content, researchers have also collected datasets of content known to be the output of generative models~\cite{RCV18}.

While it may be possible to design an effective detector against the current generation of deepfake generators, in the long-run this is likely to be a losing battle or at best a stalemate~\cite{G17}. Indeed, detection in ML is even more likely to result in an arms race than traditional system and network anomaly detection. This is because ML algorithms developed to create synthetic content, such as GANs, involve by design a generator trained to evade detection. This means that as the detector gets better to tell apart synthetic from natural content, so can the generator that creates synthetic content. Every time a new detector is deployed, the generator can be re-trained to evade the new detector. Improving the generator is efficient as long as enough information about the discriminator can be quickly gathered. At every iteration the generator gets better, as does the detector, but this process may never converge to a setting that steadily favors the defender if retraining the generator continues to be less costly than coming up with an improved discriminator in this setting. For example, a detector that uses blinking to identify deepfakes~\cite{LCL18} could be effective on the short term, but eventually it will likely drive the development of deep fake generators that correctly emulate human eye blinking. The end result is that sites hosting videos cannot rely on the resulting detector to robustly identify fake videos in the long term, at least not solely by analyzing the video itself. To summarize, progress in generative model research is likely to continue to give an edge to those creating fake content in the long term. In the short term, this would very much resemble the status quo in signature-based malware detection where defenders are constantly defining signatures for new forms of malware.

\subsubsection*{Content provenance} The second approach seeks to improve the provenance of human forms of digital communication. By provenance, we mean building a secure record of all entities and systems that manipulate a particular piece of content~\cite{MHB06}.  Consider our deepfake example mentioned above. The goal of a data provenance approach would be to identify deepfakes as content that was digitally synthesized instead of being captured using a camera. A fairly obvious solution would be to equipe every digital camera with a tamper proof cryptographic content signing key. The camera will use the key to sign all video clips that it exports. This way, every video clip is accompanied by a digital signature that identifies the physical camera on which the clip was shot. Such functionality is already available with applications like Guardian Project's ProofMode. Presumably, a deepfake generator would be unable to sign a fake video because it does not have the signing key embedded in the hardware of a real video camera. However, due to key creation, distribution, authentication, and other issues, implementing this in practice would be logistically difficult. This would be further complicated if the content is post-processed by users (e.g., to crop or apply filters) because the edits will have to preserve the image or video signature~\cite{NT16}. Alternatively, one may be able to defeat the provenance system using the ``analog hole'' attack---simply play an unsigned deepfake video on a screen, and record the screen using an approved camera that properly signs videos. In the absence of other identifying factors like the physical location in the content verified by the signature, it would most likely be difficult to detect this attack from artifacts added by the screen, as discussed previously.  

\subsubsection*{Total accountability} The third technological defense is a regime of total accountability. Consider a fictional public figure, Bernie, who is concerned about fake videos, such as deepfakes, and is willing to take extreme measures to protect himself.  In principle, Bernie could record every minute of his life on a tamper proof camera that signs and timestamps all of its captured videos. If a deepfake of Bernie is published, Bernie could prove that at the purported time that the deepfake took place, he was engaged in a slightly different activity. This after-the-fact defense would not mitigate the potential damage to Bernie's public image, but would let Bernie prove that the deepfake is a forgery. Of course, this cure may be worse than the disease: the potential loss of privacy from this 24/7 self-surveillance may cause more harm than the concern over deepfakes. In truth, we are fairly certain that a different instantiation of total accountability is required to avoid having every person create their own version of the ``Truman Show''. 

Such an approach is in fact being explored within industry: a product called Amber Authenticate proposes to have cameras periodically compute video signatures and record them publicly on a blockchain. This does not require sharing the actual content of the videos. Nevertheless, it is then possible for anyone to access the hashes recorded on the blockchain and, given video footage, verify that the hash of this video footage corresponds to what was recorded on the blockchain. This allows one to authenticate the video as having been recorded by the camera at a certain date and time---as claimed by the author of the video.

\section*{Why is emulated human behavior a problem that is challenging to solve purely through technology?}

It is clear from this discussion that technology alone cannot address the challenges of fake content emulating human behavior through machine learning. Furthermore, the very involvement of humans sets out the potential limitations of any solution addressing this problem purely through technology. Take the example of fake news. Research has shown that humans actively seek to reinforce their opinions~\cite{BMA15}. People will want to hear what they like to hear. As a consequence, each one of us reinforces their bubble of opinions through the selectivity and bias of our online connections. This effect is more prevalent in certain demographics, age being the demographic characteristic observed in a recent study to have the most significant effect on sharing fake news with their online connections~\cite{GNT19}. Even if individuals attempt to fact check their opinions and break out of a bubble of opinion, finding unbiased information can be difficult~\cite{T18} and have a limited effect on their misperception~\cite{NR10}.

This problem is not new but it is aggravated by the scale and level of personalization afforded by generative ML. Combined with large intentional or unintentional leaks of private data describing traits of our personal preferences and personalities, these techniques are instrumental in generating media that can effectively manipulate populations.

Does this mean we need to give up some forms of anonymity? Social networks have already started providing mechanisms to distinguish anonymous users from users linked to real-world individuals. For instance, Twitter adds a blue star next to users whose identity has been verified. This verification is optional but recommended by Twitter for populations often targeted by bots, such as journalists or celebrities. Obviously, this will not solve the bubbles of opinions. However, it could begin to address the crisis of content authenticity that will result from increasingly high exposure of Internet users to content generated by machines. However, this is not a silver bullet. This loss of anonymity, along with approaches for improving content provenance, may have unintended negative consequences: e.g., by enabling the tracking or identification of dissidents, minorities, or otherwise vulnerable groups that might face repercussions for content they created.

\textbf{The key is however developing public policy, legal, and normative frameworks for managing the malicious applications of technology in conjunction with efforts to refine it.} Law and policy typically lag behind technological innovation, because the implications of new technologies and how to address them can take time to come into focus and/or emerge as politically salient enough for policy makers to pay attention to them. Asimov summarizes this well: ``Science gathers knowledge faster than society gathers wisdom.'' For the most part, this lag is a feature of a vibrant innovation ecosystem, because it enables experimentation, risk-taking, and freer exchanges of ideas and capital. It can emerge as a bug, however, when innovation results in rapid paradigm shifts in the relative symmetry between malign offensive uses of technology and technological efforts to defend against them---as the case was for information technology, and potentially now for ML. New technologies are not always ``penetration tested'' from a policy perspective, because the forces behind innovation often focus on the positive applications and are not always incentivized to think proactively about malicious applications, which are easily cast aside as ``somebody else's problem, not mine.''

Here today, we see machine learning as creating numerous policy challenges, some of which mirror many of the past (and current) concerns, and others that are more novel and may create opportunities. For example, what recourse does a victim have when a fake video is created of them? Individual companies already have complementary processes in place to help users report and remove content from their platforms, but how would they be applied to content created via machine learning? What rights or options, if any, do individuals have, as a practical matter, for protecting their content from being consumed for the purposes of future content being faked? What is the appropriate response against governments who deploy the technology to interfere with liberal democratic institutions? Even the basic definition of content ownership becomes murky in this new reality: are the data and algorithms created by machines the property of individuals who designed and built these machines or a collective holding of the individuals who contributed the data that the machine was learned with? Answers to these questions and many, many more are going to shape our society and future. We simply cannot wait to see the harms emerge before dealing with for them.

We argue for thinking comprehensively about the toolkit for dealing with these and other potential harms, and for disaggregating a given problem into as many smaller pieces as possible. For example, in the case of manipulated videos such as deepfakes, one approach to disaggregating the problem is to organize our thinking around the different actors with a stake in the matter, and identifying policy tools aimed at nudging, shaping, or informing their behavior. From this perspective, there are a number of different actors whose behavior and decisions are relevant to fake content. These include (among others) the authors of fake content; authors of applications used to create fake content; owners of platforms that host fake content software; educators who train engineers in sensitive technologies; manufacturers and authors who create platforms and applications for capturing content (e.g., cameras); owners of data repositories used to train generators; unwitting persons depicted in fake content such as images or deepfakes; platforms that host and/or distribute fake content; audiences who encounter fake content; journalists who report on fake content; and so on.

Breaking down the problem in this way allows us to think more creatively about the range of policy tools relevant to the task at hand, puts us in a stronger position to identify the right policy tool(s) for the job of shaping the behavior of a given actor and, if and when necessary, develop new tools. As is the case for research on the security of computer systems, a precise threat model capturing the goal and capabilities of actors relevant to the system being analyzed is the first step towards principled defenses. Ultimately, this approach has the potential to coalesce into a more comprehensive strategy that aligns incentives across the different actors. No single tool may prove decisive, but a comprehensive approach that draws on multiple tools affecting different actors could materially move the needle.

For instance, certain groups of would-be authors of deepfakes---politicians, for example---could commit to not depicting their rivals in deepfakes. In a democracy such as the United States, we submit that many (and perhaps even most) politicians would likely find a norm along these lines attractive, under a mutual assured destruction sort of logic: all things equal, most politicians in a democracy would prefer to operate and campaign in a world where they and their opponents do not resort to outright fabrications, as opposed to one where such behavior is accepted. The U.S. Congress or a state legislature could endorse a norm along these lines in a joint resolution, and the legislative campaign committees could do so as well, and even withholding funding for candidates that violate the norm. Obviously, some politicians and other political actors might reject or violate such a norm. This would be even more concerning in less democratic societies where totalitarian governments may themselves be using ML to enable propaganda at scale. Platforms could help by rejecting fake content along these lines when they discover them, or at least downgrade them in their promotion algorithms. Each of these measures, on its own, is incomplete, but together they could have impact.

Of course, politics isn't the only domain where fake content could gain traction. Bullies and extortionists will also find uses for the technology. Here as well, an actor-centric approach yields many possibilities. For example, legislatures (or courts) could clarify that depicting a third person in a deepfake without their consent is defamation; victims would then have a cause of action for recovering money damages from authors. Legislatures could also establish criminal penalties, along the lines of legislation pending in the California legislature. Some malicious authors will hide their identities or may not have deep pockets, so holding them liable is only a partial solution. Technical measures may be useful in this context, despite their limitations. Indeed, authors of software capable of producing deepfakes could be incentivized to include cryptographic signatures to aid detection of deepfakes, perhaps by holding developers who do not include a signature liable for works created using their software. App stores and other fora for acquiring software could refuse to carry software that lacks this capability, and be incentivized to do so through mandates in law or through civil liability. Obviously, software that lacks such a capability will still be available elsewhere, but these or other barriers could deter casual users while limiting the options available to power users with malicious intentions. 

Meanwhile, platforms that host content could be required to not only establish a procedure for receiving complaints about deepfakes, as some have done voluntarily already for content that violates their terms of service or community standards, but to also provide a concise overview of the principles behind such standards. The Federal Trade Commission could then hold the platforms accountable to these published commitments using its unfair trade practices authority. Platforms could also label content known or suspected to be machine generated. Obviously, the platforms could not label all machine generated content as such for the technical reasons described above having to do with false negatives and positives~\cite{BBC+19}. This balance will be more difficult to achieve as many photo or video editing tools are likely to start including some amount of machine learning. Nevertheless, these labels could still be useful in situations where the authenticity of the content is of paramount importance to its authors and viewers. User research will be useful here to find an implementation that best ensures effective long-term user interaction with such labels and avoids pitfalls such as the implied truth effect on unlabeled content~\cite{PBC+19}. Educators who train the next generation of engineers could elevate policy and ethical literacy as important facets of technical education. And bolstering digital media literacy, especially for demographics at highest risk of being deceived, is also essential. Indeed, research has found that correcting misperceptions through the presentation of factual evidence has limited effect, and can sometimes be counterproductive by strengthening such misperceptions~\cite{NR10}.

We present these governance interventions to illustrate how breaking the problem down in this way can yield insights into the possibilities for shaping behavior. None of these governance interventions is a silver bullet, in the same way that the technical possibilities described above are not. Some of them also raise other challenges or concerns, and implicate difficult tradeoffs across other important values or equities. In addition, determined bad actors will often find ways around them. But for less determined bad actors, interventions along the lines of what we describe above from the technical and policy perspectives could prove decisive if put in place jointly. And having to operate in this governed  environment would make it costlier for even the determined bad actors to create and spread malicious content.

\section*{Conclusion}

As Turing envisioned in 1950, machines are on track to become well capable of producing any form of human communication. It is likely that they will eventually simulate fake human behavior effectively and allow for the creation of sophisbots. Perhaps one of the most pressing technical questions for the first half of this century is thus how we distinguish reality from the synthetic in our evolving world of thinking machines. Answers to this question will shape how we as a broader society communicate and live long into the future. In that regard, the test that Turing envisioned in 1950 is more relevant than ever: will humans continue to be able to identify sophisbots, albeit using increasingly higher levels of knowledge and logic abstractions, until we are able to create an artificial intelligence? The call is clear. Let us as a technical community commit ourselves to embracing and addressing these challenges as readily as we do the fascinating and exciting new uses of intelligent systems.

\appendices

\ifCLASSOPTIONcaptionsoff
  \newpage
\fi

\end{document}